\def\cm{{\rm\thinspace cm}}

\def\erg{{\rm\thinspace erg}}

\def\keV{{\rm\thinspace keV}}

\def\Msun{\hbox{$\rm\thinspace M_{\odot}$}}

\def\s{{\rm\thinspace s}}

\def\pcmcu{\hbox{$\cm^{-3}\,$}}

\def\spose#1{\hbox to 0pt{#1\hss}}
\def\approxlt{\mathrel{\spose{\lower 3pt\hbox{$\sim$}}
        \raise 2.0pt\hbox{$<$}}}
\def\approxgt{\mathrel{\spose{\lower 3pt\hbox{$\sim$}}
        \raise 2.0pt\hbox{$>$}}}

\documentstyle[psfig]{mn}

\title
[Spectral States of black hole Candidates]
{Magnetic flares in accretion disc coronae and the Spectral
States of black hole candidates: the case of GX 339-4.}

\author[T.~Di Matteo, A.~Celotti and A.C.~Fabian]
{T.~Di Matteo $^1$, A.~ Celotti $^{1,2}$ and A.C.~ Fabian $^1$\\
{$^1$Institute of Astronomy, Madingley Road, Cambridge, CB3 OHA}\\
$^2$ S.I.S.S.A., via Beirut 2--4, 34014 Trieste, Italy\\}

\begin{document}

\maketitle

\begin{abstract} 
We present a model for the different X-ray spectral states displayed
by Galactic Black Hole Candidates (GBHC). We discuss the physical and
spectral implications for a magnetically structured corona in which
magnetic flares result from reconnection of flux tubes rising from the
accretion disk by the magnetic buoyancy (Parker) instability. Using
the observations of one of the best studied examples, GX339-4, we
identify the geometry and the physical conditions characterizing each
of these states. We find that, in the Soft state, flaring occurs at
small scale heights above the accretion disk. The soft thermal--like
spectrum, characteristic of this state, is the result of heating and
consequent re-radiation of the hard X-rays produced by such
flares. The hard tail is produced by Comptonization of the soft field
radiation. Conversely, the hard state is the result of a phase in
which flares are triggered high above the underlying accretion disk
and produce X-rays via Comptonization of either internal
cyclo--synchrotron radiation or of soft disk photons.  The spectral
characteristics of the different states are naturally accounted for by
the choice of geometry: when flares are triggered high above the disk
the system is  photon--starved, hence the hard Comptonized
spectrum of the hard state.  Intense flaring close to the disk greatly
enhanced the soft--photon field with the result that the spectrum
softens.  We interpret the two states as being related to two
different phases of magnetic energy dissipation. In the Soft state,
Parker instability in the disk favours the emergence of large numbers
of relatively low magnetic field flux tubes. In the hard state, only
intense magnetic fields become buoyant and magnetic loops are able to
rise and expand in the coronal atmosphere. The model can also
qualitatively account for the observed short timescale variability and
the characteristics of the X-ray reflected component of the hard
state.

\end{abstract}


\section{Introduction}
Galactic X-ray sources are classified as black holes candidates (GBHC)
if either the measured binary mass function 
indicates the presence of an object with $M \approxgt 3\Msun$ (for a
review see Tanaka \& Lewin 1995) or their high energy X-ray spectra
and temporal variability are similar to other GBHC.

For many years GBHC have been known to radiate in five different
spectral states, defined by the observed spectral components
and flux level typically in the $1-10 \keV$ band. Systems in the {\it 
Hard/Low state} emit most of their energy in a hard tail which can be
represented as a power law with a photon index $\Gamma \sim 1.3-1.7$
and an exponential cut--off at about 100 \keV (Tanaka \& Lewin 1995,
Zdziarski et al. 1997). Most of these objects are also observed in a
{\it Soft/High State}, when most of the energy is emitted in a
blackbody component with characteristic temperature in the range of
$0.6- 1 \keV$. In addition to this thermal component the spectrum
comprises a power law, characterized by a slope $\Gamma=2.0-2.5$,
which dominates above $\sim$ few keV.  It was originally thought that
the GBHC systems were much more luminous in the soft than in the hard
state, but recent observations of GX339-4 and Cyg X-1 have shown that
despite their dramatic spectral changes the bolometric luminosity 
changes only slightly.

Although the hard and soft states are the most common ones,
occasionally GBHC have been observed in three other states (these
occur as transitional events and have been observed rather
infrequently).  The {\it Intermediate State} is seen during
transitions between the low and high state. In the {\it Very High
State}, GBHC have the highest luminosity: the high-energy power law
component has a flux comparable to the soft blackbody one and the high
energy emission does not show any sign of a cut-off.  In the {\it Off
State}, spectra are completely dominated by a power-law ($\Gamma=1.7$)
component with a flux level lower than in the hard state by a few
orders of magnitude.

Another important signature of the different source behaviour in the
different spectral states is the variability pattern.  In the soft
state the normalization of the power law has been observed to vary
whereas the soft blackbody component is very stable.  The emission in
the hard state instead shows extreme variability on time scales as
short as $10^{-3}$ s. Short timescale variability is also
observed in the power law of the very high state.

Although temporal and spectral behaviors of GBHC have been widely
studied in the last few years, the nature of the different spectral
states and in particular the mechanism driving the transition from one
to the other are largely unknown. The thermal radiation,
characteristic of the high, very high and intermediate states is
generally modeled to be blackbody emission from a (standard) optically
thick accretion disk.  The power law component is generally
attributed to inverse Compton radiation from an optically thin corona
sandwiching the accretion disk.  The presence of such a hot medium was
initially hypothesized in the context of Seyfert galaxies (e.g. Haardt
\& Maraschi 1993) and because of the similarity between the spectra of
Seyfert galaxies and those of GBHC in the hard state, the same model
was naturally extended to the galactic objects too.  In the Seyferts
scenario it has also become necessary, in order to account for the
different ratios of UV vs X-ray luminosity and the extremely short
variability time scales observed, to assume that such a corona actually 
consists of localized active regions (e.g. Haardt, Maraschi \&
Ghisellini 1994, Stern et al. 1995, Nayakshin \& Melia 1997a,b). Such
regions could be the end result of impulsive magnetic energy
dissipation (i.e. reconnection) in flux tubes emerging from the disk
by buoyant instability (as originally suggested by Galeev, Rosner \&
Vaiana 1979).

In GBHC different geometries for the spatial distributions of the hot
and cold matter associated with the accretion flow have been proposed.
Some models (Dove et al. 1997, Gierlinsky et al. 1997, Poutanen,
Krolik \& Ryde 1997) deduce the geometry of the emitting regions from
the spectral analysis of the different states.  In particular, the
authors of such models argue that because the reflection component in
the hard state of GBHC is much less prominent then in Seyferts, the
inner radius of the disk must be far away from the black hole, in
order to subtend a small solid angle to the X-ray emitting
region. Most of the energy would be dissipated in a thermally hot
central cloud/corona--like structure which, by Comptonization,
produces the observed spectrum.  In the soft state the optically thick
cool disk is postulated to move inwards and the majority of the
dissipated energy would emerge in the form of a blackbody-like
spectrum.  In these scenarios, though, no physical mechanism is
provided to explain the origin of such drastic changes in the geometry
of the inner region between the two states.  A possible interpretation
has been discussed by Esin, McClintock and Narayan (1997) in the
context of advection-dominated accretion flow (ADAF) solutions. In
these models the inner compton cloud in the hard state is identified
with an advection dominated zone (with still an outer thin disk at
larger radii). As the accretion rate increases an ADAF is no longer
allowed and it shrinks in size. The outer thin disk moves inwards and
the spectrum changes form hard to soft.  It should be noticed that
although the geometry of the X-ray emitting region in both classes of
models (the Compton cloud and the advection-dominated) predicts the
observed lack of reflected emission, it can not explain the fast
variability time scales observed in the hard states of GBHC.

In this paper we show that the different spectral components and the
different spectral states of GBHC can be easily reproduced within the
context of a magnetically structured corona by considering flares
triggered at different scale heights above an accretion disk. We
expect such a corona to form because the strong magnetic fields,
continuously generated by the dynamo action in an accretion disk, are
strongly buoyant and are forced to invade the region sandwiching the
disk itself.  Once outside the disk the magnetic flux tubes can
reconnect efficiently and dissipate part of the accretion energy in
localized active flares.


Here, we shall, at first, constrain the geometry and the energy
dissipation distribution of the emitting regions in the different
states exclusively from the detailed broad band spectral information
provided by the observations.  In Section 2 we will give an overview
of our model, and specify the parameters which can be derived almost
directly from observations and those related to the system geometry.
In particular, we will consider the spectral implications of having
localized active regions at different scale heights above the
accretion disk.  In section 4 we will obtain the parameter spaces
which characterize the different spectral states by applying our model
to GX339-4 (we summarize the data collected from the literature in
section 3).  In section 5 we will relate the properties of the active
coronal regions derived from the physics of radiation, to those of a
magnetically structured corona. We will show that our model is
completely consistent with the idea that X-ray flares are the after
effect of reconnection of magnetic flux tubes rising from the
accretion disk due to magnetic buoyancy. The different spectral states
can then be identified with different phases of flare activity. Such
phases would, in turn, be regulated by the physical conditions in the
accretion disk relevant for the onset of buoyancy instability.
Finally, in Section 6 we briefly discuss our model predictions for the
variability and the reflection properties in the different states.

\section{The model}

\subsection{Emission from a magnetically structured corona}
As already mentioned, we adopt a model for a magnetically structured
accretion disk corona in which localized active regions (magnetic
loops) form as a result of magnetic buoyancy instability in the
underlying accretion disk.  Buoyancy constitutes a mechanism that
channels part of the energy released in the accretion process directly
to the corona outside the disk.  Therefore, following Di Matteo,
Celotti \& Fabian (1997), we assume that a significant fraction $f$ of
the accretion power, $\eta L_{\rm Edd}$, is dissipated magnetically in
the corona while the remaining fraction $(1-f)$ is dissipated
internally to the disk.  The magnetic field strength $B$ in the
magnetic flux tubes is therefore given by
\begin{equation} 
B=9\times10^7\left(\frac{f\eta}{N_{\rm tot}}\right)^{1/2}
r^{-1}\left(\frac{10\Msun}{M}\right)^{1/2}\;\; \;\;{\rm G}, 
\end{equation}
where $N_{\rm tot}$ is the number of magnetic loops emerging from the
disk, $L_{\rm Edd}$ is the Eddington luminosity, $r$ is their typical
size in units of the Schwarzschild radius, $R_{\rm s}$, and $M$ is the
black hole mass which we take $M=10\Msun$.  Although a number $N_{\rm
tot}$ of magnetic loops become buoyant, there will only be a certain
number $N$ of loops reconnecting (i.e. producing flares) at
any given time.  We take $N=10$ as the number of active regions needed
to explain (from simple Poissonian noise) X-ray fluctuations of a
factor $\sim 2$ on short timescales which are typically observed in
GBHC.

Here we consider a situation in which the majority of the
accretion power, say $f\sim$ 80 per cent, is released into the
magnetic structures in the corona and only $\sim$ 20 per cent in the
optically thick disk.  Such an assumption implies that the X-ray
emission from the active coronal regions plays a fundamental role in
the production of the different spectral components observed in the
various states of GBHC, as we shall show in the course of this
section.

The hard X-ray emission is produced in the corona by thermal
Comptonization of soft seed photons produced both as blackbody
radiation in the underlying accretion disk and as cyclo-synchrotron by
the same hot electrons pervading the corona.  Cooling by
cyclo-synchrotron emission and its relative Comptonization are not
usually taken into account in disk--corona models, but they can be the
dominant radiative processes in the GBHC scenario (Di Matteo et
al. 1997).

The luminosity and the effective temperature of the quasi-thermal
component emitted by the accretion disk are consistently estimated as
due to both the intrinsic energy dissipation and to the re--radiation
of the coronal hard X-rays impinging on and then being
absorbed/reflected by the disk.

The relative importance of all of these different processes is mostly
and critically dependent on the assumed geometry of the system.  We
therefore considered the predictions of this schematic model (more
details are given below) against the observed spectral properties of
one of the most well known GBHC, namely GX339-4, in order to: a)
determine the dominant emission processes in the different spectral
states; b) explore the role of the spatial and geometrical
distributions of such active regions and set any possible
constraint. As we shall see in the following sections, the observed
spectral states strongly limit the geometrical (and consequently
physical) characteristics of active regions.  

\subsubsection{Model parameters}

The Comptonized (energy) spectrum is described by a power law of slope
$\alpha$ with an exponential cut-off at an energy $E_{\rm C}$.
$\alpha$ is a function of both $\tau$ and $\theta$, the Thomson
optical depth of the scattering region and the dimensionless electron temperature ($\theta= kT/m_{\rm e}c^2$),
respectively. This can be expressed as $\alpha \simeq{-{\rm ln}P}/{{\rm
ln}(1+4\theta+16\theta^2)}$ where $P$ is the average scattering
probability $P=1 + {{\rm exp}(-\tau)}/{2}\left({\tau}^{-1}-1\right) -
{2\tau}^{-1} + ({\tau}/{2})E_1(\tau)$, and $E_1$ the exponential
integral (Zdziarski et al.  1994 -- also consistent with the
analytical approximation $\alpha=0.16/(\tau\theta)$ by Pietrini \&
Krolik 1995). Two more parameters describe the soft
(disk) blackbody component, namely its effective temperature $T_{\rm
soft}$ and luminosity $L_{\rm soft}$.

Some of the above parameters can be derived almost directly from
observations. The electron temperature $\theta$ of the coronal regions
is related to the observed cut-off energy, so that $E_{\rm C}\simeq
(2-3) kT$ (depending on the value of $\tau$).  Clearly observations
also constrain the slope of the hard X-ray power law and given the
temperature and $\alpha$, $\tau$ is also determined.  Furthermore, as
it has already been pointed out, the bolometric luminosity of GBHC
only changes by small factors despite the dramatic changes in spectral
shape. This implies $L_{\rm soft}
\approx L_{\rm hard} \approx L_{\rm tot}$ 
(where the subscripts 'soft' and 'hard' refer to the soft and hard states).

As we shall see, the other parameters (notably $T_{\rm soft}$) are
related to the system geometry. We associate a temperature to the disk
blackbody due to intrinsic dissipation in the disk itself ($T_{\rm
soft}=3.8\times 10^7 [(1-f)/M]^{1/4} (r_{\rm disk}/50)^{-1/2}$ K).
However, we also consider the quasi-thermal emission, due to the
re--radiation of the hard X-rays, which is therefore a function of the
height of the active regions.  Flares triggered close to the disk
would heat up small areas of the accretion disk (small covering
factor), but contribute to the hottest blackbody emission. On the
contrary, flares at large vertical scales irradiate much larger areas
but the locally reprocessed energy density is small compared to the
intrinsic dissipated flux and thus do not affect $T_{\rm soft}$.

In order to give a simple description of such geometrical effects we
schematically represent the active regions as spherical volumes of
typical scale (radius) $r$, distributed at different scale heights $h$
above the (geometrically thin) disk, and
extending over an area out to a  radius $r_{\rm disk}\approx 50$.
Where $r$, $r_{\rm disk}$, $h$ and $R_{\rm hot}$ (see below)
are all in units of Schwarzchild radius, $R_{\rm S}=2GM/c^{2}$. 

If we then assume that half of the luminosity from each dissipation
region is directly detectable, then the amount of flux per solid
angle going back and impinging onto an area of radius $R_{\rm hot}$ of
the disk is given by
\begin{equation}
F(R_{\rm hot})dR_{\rm hot}= \frac{L}{4}\frac{ h}{(R_{\rm hot}^2+h^2)^{3/2}}dR_{\rm hot}
\end{equation}
where $L=L_{\rm tot}/N$ is the luminosity of one active region.  The
re-radiated flux from this same area (in a steady situation)
corresponds to $\sigma T_{\rm hot}^4=F(R_{\rm hot})$. For a given
height $h$, we define a typical relevant radius $R_{\rm hot}$ as the
radius at which the temperature has decreases by a factor $\sim 2$.
This corresponds to $R_{\rm hot} \approx 2.3 h$ (or approximately $=
2.3 h +r$ for small $h$).

Consistently, the soft photon field from the disk might be amplified
in those areas, of radius $R_{\rm hot}$, located below the active
regions and even dominate over the intrinsic cyclo--synchrotron
radiation. This is more likely to happen in those regions close to the
accretion disk itself (small $h/r$).

This is the key feature of the model: the active coronal regions
play a crucial role in the production of both the hard power-law and
the hot blackbody components observed in the X-ray spectra of GBHC.

The hot blackbody emission visible in the soft state is the result of
the heating, and consequent re-radiation of the hard X-rays produced
by those magnetic flares triggered at low $h$.  The hard power-law
tail in the soft state (and in the hard state if the active regions
are large enough), is due to inverse Compton scattering of the soft
disk photons by the high energy electrons.  In the hard state however,
the hard power-law component is usually due to inverse Compton of the
photons produced as thermal cyclo--synchrotron radiation. From the
above discussion, it is then apparent that the only free parameters in
this scenario are the typical scales, height $h$ and dimension $r$, of
the flares above the disk.

In the next two sections we shall directly investigate the spectral
implications of having flares at different heights $h$ by comparing
the model predictions with the observations of GX339-4. 

\section{The GX339-4 data}

GX339-4 shows a wide range of energy spectral shapes and variability
patterns, allowing it to be classified as being in the off, low,
intermediate, high, or very high states.  In order to examine these
different states we have considered the observations available in the
literature on the full-band spectra, which covered as many states as
possible.  Here we review the data for each one of the spectral states
of GX339-4, which will be represented by their spectral model fits in
our figures. These are not intended to represent any realistic
physical model but merely provide a way of describing the data.

\subsection{Soft/High State}

Observations of GX339-4 in its soft state have been reported by
Makishima et al. (1986). The object was observed simultaneously in the
X-ray and optical bands, using the X-ray satellite {\it Tenma} ($2-20
\keV$) ~and the Anglo Australian Telescope, 
respectively.  The X-ray spectrum comprises a very soft component
modeled with a disk-blackbody at a temperature of $0.7 \keV$ and a
hard tail represented by a power law. We have selected two of the
reported observations which show an increase in the intensity of the
hard tail (the soft X-ray component is always very
stable). Observations by Ariel-6 ($2-10 \keV$) of this soft component
are also presented by Ricketts (1983) and the spectrum is described
with a power law of photon index $\Gamma=4.5$. GX339-4 was observed by
the HEAO 1 ($12-200\keV$) satellite (Nolan et al. 1982) and these
data, also, showed these two spectral components. The two
observations, fall 1977 and spring 1978, were fitted using model
functions which are the sum of a power law and bremsstrahlung
emission.

At higher energies, the hard tail of GX339-4 has been observed by
BATSE (20-300 \keV; ~Harmon 1994) and SIGMA (35-1300 \keV; ~Bouchet et
al. 1993). In 1991 both instruments simultaneously observed the source
during an outburst making a transition from the hard to the soft
state. All the spectra can be well represented by a thermal
bremsstrahlung model ($kT \sim 100 \keV$).

\noindent{\it Optical data:} 
Doxey et al. (1979) and Motch, Ilovaoisky \& Chevalier (1985) observed
the object in May--June 1978 and Feb--Mar 1982, respectively; its
magnitude was reported to be V=16.6-16.7, similar to the AAT
observation (V=16.5) reported by Makishima et al. (1986).

As already mentioned, in the soft X-ray state the source does not show
either optical or X-ray short time scale variability.

\subsection{Very High State}

GX339-4 was observed with {\it Ginga} (4--28 keV) 
in its very high state (Miyamoto
et al. 1991), with a spectrum similar to the hard state, 
consisting of a low energy component (disk blackbody at $kT \sim 1
\keV$) and high energy tail (power-law with $\Gamma=2.5$).
The X-ray intensity was a factor 2-3 larger than in the high/soft
state, but unexpectedly showed very rapid variations on time scales of
less than several minutes in the power law tail.

\subsection{Intermediate State}

Data from the EXOSAT ME (1--20\keV) ~has revealed the presence of an
intermediate state, characterized by a soft spectral
component (fitted by a blackbody with $kT=0.42 \keV$) and a relatively
steep, harder power-law tail ($\Gamma=3.5$; Mendez \& van der Klis
1997).

\subsection{Hard/low State}

GX339-4 was observed in its hard state on a few occasions with {\it
Ginga} in Nov 1991 (Ueda, Ebisawa \& Done 1994). In
all cases the energy spectra were well represented by a single power
law component with a typical photon index of $\Gamma \sim 1.7$.  
Ariel--6 observations (May 1981) found the source in its
hard state with a spectral slope of $\Gamma=1.5$ in the $1-20\keV$
band (Ricketts 1983; Motch et al. 1983).

The high energy X-ray spectrometer on board of OSO 8 observed GX339-4
in Sep 1976. Dolan et al. (1987) fitted the hard state spectrum
with a power law with $\Gamma\sim 2$ in the $16-151 \keV$ energy
range.  BATSE observed GX339-4 in its low state in Jul--Sep
1991 (Harmon et al. 1994) and the $20-300 \keV$ spectrum during this
outburst was fitted by a bremsstrahlung model ($kT \sim 100
\keV$ for the July observation and $kT \sim 60 \keV$ at the end of the
outburst in September).  Simultaneous OSSE ($59-400 \keV$) and SIGMA
($35-1300 \keV$) data (Sep 1991) were reported by Bouchet et
al. (1991) and Grabelsky et al. (1995), respectively. We include
bremsstrahlung fits to these data also.

\noindent{\it Optical data:} The Ariel--observations (May 1981) were
simultaneous with optical fast photometry (Motch et al. 1981; Motch et
al. 1983).  During this period the optical counterpart was reported to
be extremely bright, with V=15.4.  More remarkably, the optical flux
showed very short time scale activity (flares as short as 10--20 ms
during which the flux increased by a factor up to 5; Motch et
al. 1983).  The source also appeared to be redder in the infrared band
during the May 1981 activity phase (Motch et al. 1982).

Note that short timescale variability has consistently been observed
when the object is in its hard state.

\subsection{Off state}

Ueda at al. (1994) reported on observations with {\it GINGA} from 1989
to 1991. The 1989 data were also (see above) well represented
by a power law ($\Gamma \sim 1.7$) but with much lower intensity
than in 1991.  During the off state observed by EXOSAT in Apr 
1985 the source was found to exhibit a hard power law
spectrum ($\Gamma=1.8$; Ilovaisky et al. 1986). The spectrum
of the off state was also recorded in Sep 1988 by HEXE ($30-180
\keV$) and fitted with a power law of 2.2 (Dobereiner et al. 1989).

In its off state the optical counterpart of GX339-4 becomes
exceptionally faint, $V \approxgt 18$ (Motch et al. 1983).

\section{Application to GX339-4}    

We can now compare more specifically the model predictions against 
the data available for GX339-4. 

We recall that the main parameters that allow us to describe the
spectrum are determined by our choice of geometry, i.e. $(h,r)$.  The
blackbody emission is calculated self--consistently for a given
height: the intrinsic energy dissipation in the accretion disk gives
$T_{\rm soft}$; the re--radiation of the hard X--rays determines the
effective temperature $T_{\rm hot}$ in those areas, of dimension
$R_{\rm hot}$, of the disk heated by the active regions.

In the regime considered, the cyclo--synchrotron emission is
self--absorbed \footnote{The formalism used to determine the
self--absorbed frequency is described in Di Matteo et al. (1997).}.
The soft photons (both in the effective blackbody component and in the
cyclo synchrotron peak) are Comptonized in the active regions.

It should be stressed that, as our purpose is to determine any
constraint on the global properties of the active regions, there is no
attempt to perform actual best fits to the data, but rather to define
the model parameter space consistent with the `average' of different
observations for each one of the states.

\begin{figure}
\centerline{\psfig{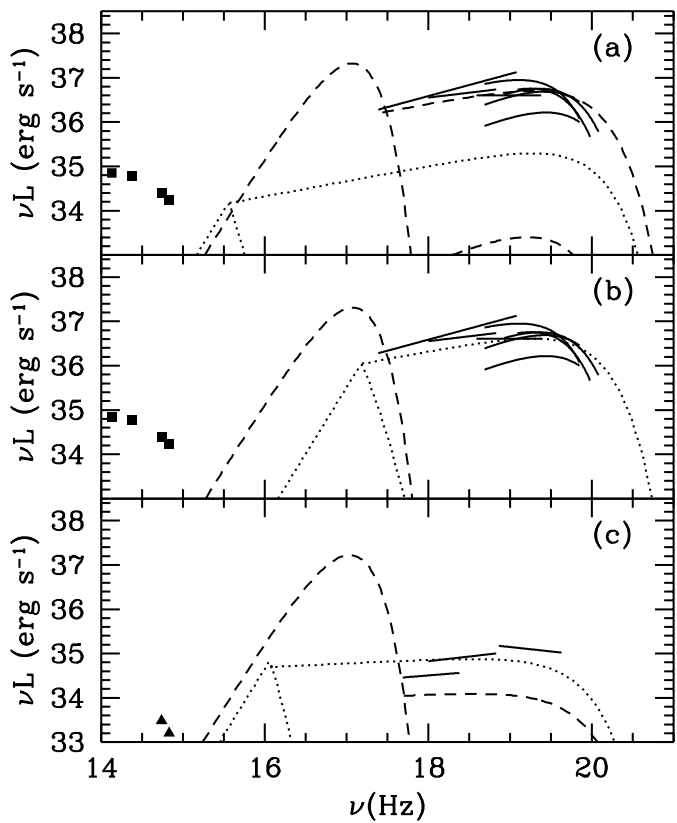}}
\caption{Representative model predictions (dotted and dashed lines) 
plotted against fits to the data (represented by the solid lines) for
the {\it hard state} (a and b) and {\it off state} (c) of GX339-4.
The flares are triggered high above the disk so that their emission does
not affect the intrinsic temperature of the accretion disk, whose
blackbody emission is represented by the dashed line.  (a) The
spectrum results from the intrinsic blackbody emission from the
accretion disk and its Comptonization, for $\tau=1.35$, $\theta=0.18$,
$r= 2$ and $h=3.3$.  (b) The hard X--rays are dominated by
Comptonization of soft seed photons originating as thermal
cyclo--synchrotron radiation in the active region (dotted lines).
$\theta$, $\tau$ and $h$ are the same as in case (a).  The dimension
of the active regions is instead $r=0.1$. (c) In the {\it off state}
the parameters are the same as those for the hard state of case (b)
apart from $\theta$ which is in this case 0.13.  The data (solid
lines) are those described in sections (3.4) and (3.5).}
\end{figure}

In the {\it hard state} of GX339-4 we have $\theta\simeq 0.18$
(derived from the spectral cut-off in the data) and $\alpha \approx
0.4-0.5$ (considering several observations) which imply $\tau=1.35$.
The soft component is not observed in this state;
this immediately implies that flares
are unable to heat up the intrinsic blackbody disk emission (which
peaks in the UV -- see Fig.~1) because they are either intrinsically
weak or located high above the disk. As we shall discuss later, the
power released in a single flare is constrained to be quite high, and
therefore we can only reproduce the hard state assuming that $h
\approx 3.3$, i.e. $\sim 10^7 \cm$ for a $10\Msun$ black hole. 

Depending on the dimension of the active regions, the high energy
component is dominated by Comptonization of either internal
cyclo--synchrotron photons or blackbody ones. This implies that $r$ is
not tightly constrained by observations, and can range between
$0.05-3.3 R_{\rm S}$: in compact (i.e. high $B$) regions the intrinsic
cyclo--synchrotron soft photon field dominates (Fig.~1b), while for
larger $r$, the photon energy density is dominated by the soft disk
photons (Fig.~1a).

Note that the two different Comptonized power--law components are
renormalised according to the relative timescales associated with the
corresponding radiative processes.  This implies that, although in
some cases the cyclo--synchrotron emission does not contribute
significantly to the observed spectrum, it still influences the
normalization of the component due to inverse Compton on soft disk
photons.


The {\it off-state} can be `fitted' with very similar parameters,
$h=3.3$, $r=0.3-1$ and a slightly lower electron temperature,
$\theta=0.13$ (the data do not provide a clear indication of a spectral
cut--off in this state).  Even a small decrease in the electron
temperature implies much less cyclo--synchrotron emission (Fig.~1c),
requiring $r$ to be adjusted quite finely in order to account for the
Comptonized hard X--ray spectrum.

\begin{figure}
\centerline{\psfig{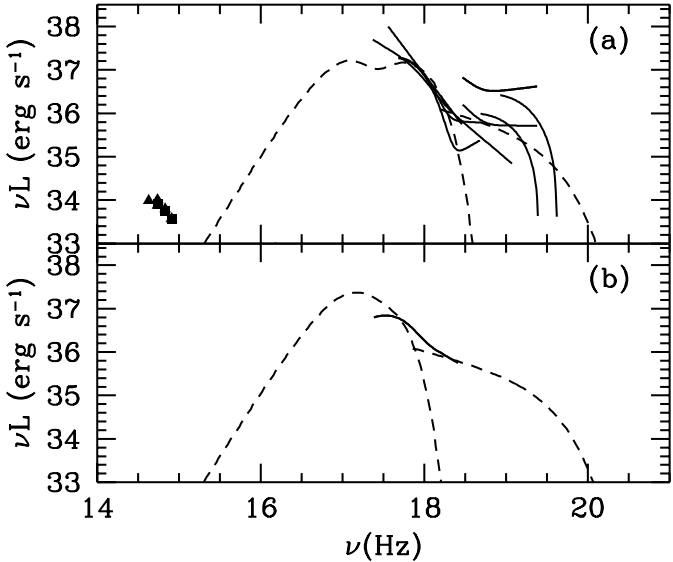}}
\caption{Model predictions (dashed lines) plotted
against fits to the data (solid lines) for the (a) {\it soft state},
and (b) {\it intermediate state} of GX339-4.  The temperature in the
disk areas of radius $R$ heated by the flares increases, giving raise
to the hotter blackbody component.  The power--law component is due to
Comptonization of these same blackbody photons.  (a) $\theta=0.08$ and
$\tau=1.35$. The dimension of the active regions is $r=0.4$ and its
height above the disk is roughly equivalent to its dimension, $h=0.1$.
(b) $\theta=0.08$, $\tau=1.35$ (as before) $h=2$ and $r=1$. The data are those described in sections (3.1) and (3.2).}
\end{figure}

The emission in the {\it soft state} is dominated by the soft thermal
component. Within the context of our model this hotter
quasi--blackbody component can be generated by the hard X--ray
radiation impinging onto and being reprocessed in the disk.  When
flares are triggered close enough to the disk, namely at $h=0.1$,
i.e. $\sim 3\times 10^5 \cm$, the temperature in the areas of radius
$R_{\rm hot}$ achieves the observationally required values to explain
the $\approx 1\keV$ blackbody emission.  Note that, if the dissipation
in magnetic flares is comparable in the hard and soft states, this
mechanism naturally accounts for the similar luminosity observed in
the two states.  Fig.~2 shows this additional blackbody component due
to flare reradiation.

The energy of the spectral cut-off implies electron temperatures lower
than in the hard state, i.e. $\theta=0.08$, and by imposing $\alpha
\approx 1.5$, the optical depth $\tau= 1.35$ is consistent with being
constant between the two main spectral states.

The dimension of the active region is constrained in a rather small
interval, $r=1-4$, overlapping with the upper range of that of the
hard state case. Because of the lower $\theta$, the
self--absorbed cyclo-synchrotron emission ($\nu L_{\nu}({\rm synchro})
\propto \theta^7$) cannot significantly contribute to the seed photons 
field for the hard tail, which is instead due to inverse Compton of
soft disk photons (Fig.~2a).

The {\it intermediate state} is also dominated by the blackbody
component but this is at slightly lower temperatures than in the soft
state.  Within our model, this most naturally arises from flares at
intermediate heights, with $h \sim 2$ and $r \sim 1$ (Fig.~2b).
Finally, in the {\it very high state} the blackbody component of GX
339-4 is similar to that observed during the soft state but the
normalization of the power--law tail is higher. This can be easily
explained by a larger covering fraction ($r\sim $ a few) of the active
regions.

In summary, the hard state requires parameters of the active regions
quite similar to those of the soft state. The two primary differences
are to do with the fact that in the soft state active regions are
mostly at small $h$ and have lower electron temperature $\theta$,
whereas in the hard state are characterised by higher $h$ and $\theta$.
All the other states can easily result from intermediate values of
these parameters.

In other words, the heart of this model is the choice of the geometry.
Depending on whether the flares are high above the disk or close to it
the spectrum changes from being hard to soft respectively (and
therefore gives rise to the hard state or the soft).  When the flares
are located high above the disk (and they are few in number), the
solid angle they subtend to the soft photon field produced in the disk
is small: the system is photon starved and the Comptonized spectrum
hard, as required by observations (synchrotron radiation is highly
self--absorbed and therefore its Comtonization highly saturated also)
A photon--starved system can naturally account for the spectral
characteristics of the hard state.  Conversely, when a large number of
flares is triggered very close to the disk surface, the soft photon
field is highly enhanced giving rise to a hotter blackbody component
(due to reprocessing in the disk itself) and a very soft
Comptonization spectrum. 

\begin{figure}
\centerline{\psfig{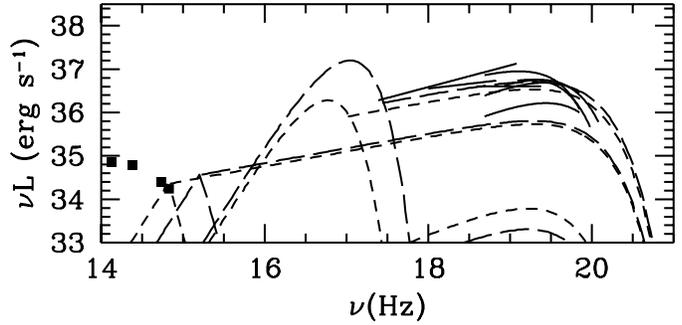}}
\caption{The cyclo--synchrotron emission peaks in the optical for 
either $f=0.9$, $N=100$ and $r=3$ (short dashed lines), or $f=0.8$,
$N=100$, $r=1.3$ (long dashed lines). $h$, $\theta$ and $\tau$ are the
same as in Fig.~1b.  This can explain the short timescale optical
variability detected in the observations by Motch et al. (1983--filled
squares).}
\end{figure}

Before concluding this section we would like to point out an
interesting feature of our model. This concerns the role of the
cyclo--synchrotron emission. If the number of active regions increases
to $N=50-100$, the cyclo-synchrotron emission peaks at much lower
frequencies and can explain the observed optical flux, as shown in
Fig.~3. An increase in $N$, in fact, corresponds to a decrease in the
intrinsic magnetic energy released in each loop. If this occurs, then
variations in optical flux are expected to occur on short timescales
(related to the size of our active regions, as indeed observed by
Motch et al. (1983) in GX339-4. This increase in covering fraction can
also explain the anti--correlation between optical and X-ray
variations (Motch et al. 1983). As we see in Fig 3, when the optical
emission is due to the intrinsic synchrotron emission (and variations
in the optical are expected) the X-rays are produced by Comptonization
of the external soft blackbody radiation (see Fig 3). As already
discussed, when this process dominates the dimension of the active
regions needs to be increased, implying a much larger covering
fraction (with $N=100$), and therefore much slower variations in the
X-rays with respect to the standard hard state in Fig 1.

\subsection{Other model parameters}

It is important to stress that the value of the other model parameters
(notably $N, f$) does not need to be tuned finely. Let us consider
them in turn.

In the hard state, an increase in number of the active regions implies
a (slow) decrease in $B$. If $N=10-50$ the Comptonization on
cyclo--synchrotron photons still dominates, but as $N$ reaches $\sim
100$ (see also Fig.~3), this component decreases significantly (as $B$
does).  This implies that the dimension of each region needs to be
increased for Comptonization of the external photons to account for
the observed flux.

In the soft state case, the most relevant modification is the amount
of flux dissipated in each region, which in turn affects the effective
temperature $T_{\rm hot}$ which decreases as $N$ increases (the active
regions are more numerous but less powerful and thus do not heat
significantly the underlying disk). For $N
\approxgt 50$ the soft blackbody component (due to re--radiation) becomes
too cool to explain observations.  

On the other hand, the power magnetically dissipated, i.e.  $f$, might
be lower. This implies that the normalization of the disk blackbody
component increases but, even for $f=0.2$, the dominant radiative
contribution in the hard state is still from cyclo--synchrotron
emission and its Comptonization. In the soft state instead the net
effect is a change in the relative normalizations of the two blackbody
components.

\subsection{Statistical fluctuations?}

We have shown that all of the different states can be obtained with
appropriate choices of $\theta$ and $h$. Clearly the fundamental point
is to interpret these findings in a coherent and self--consistent
physical picture of the whole process and in particular of the change
of the spectral states of the source.

However, before dealing with this point, a key question which needs
answering is whether the spectral information requires that in each
state the dissipating coronal region is pervaded only by active
regions with given characteristics or is consistent with a spectral
state dominating over another although `different' active regions are
present at the same time.  In other words, we examine whether it is
possible to have $N=10$ active regions associated with one particular
temperature and $h$ and say $N=7$ (a statistical fluctuation on $N$)
flares at different $h$ and $\theta$, but still have the `main' state
dominating.

We find that statistical fluctuations are inconsistent with observing
any one particular spectral state. More precisely, in order to have
the hard state, we can only have active regions at $h=3.3$ and no
regions (within a random fluctuation) at lower $h$ can be
active simultaneously.  The temperature instead is not constrained,
so that at each height $h$ regions can have different temperatures
(ranging from $\theta=0.08$ -- soft state to $\theta=0.18$ -- hard
state).  The predicted spectrum in the soft state is inconsistent with
random fluctuations for both regions at different heights and at
different temperatures.

\section{The physical picture}

In this section we discuss how the geometric representation adopted
for the different states relates to the physical picture in which
coronal loops, emerging from the disk by magnetic buoyancy, reconnect
and dissipate their energy in powerful X-ray flares.

\subsection{Magnetic buoyancy and reconnection}
From the comparison of the model predictions with the observations of
GX339-4 we find that, in order to reproduce the different spectral
states, the active regions have to be located at different heights
above the accretion disk with temperatures inversely proportional to
their height. This anti-correlation is indeed (qualitatively) expected
in the physical frame of reconnection driven by magnetic buoyancy.

Let us briefly recall the general picture associated with a
magnetically structured corona.  Strong fields in the accretion disk
are continuously generated because of the shearing and the turbulent
disk flow serves as an effective dynamo that can rapidly strengthen a
seed field.  Because intense fields are strongly buoyant they rapidly
emerge from the disk thereby releasing part of the accretion energy
into a corona in the form of magnetic energy.
This disk instability is therefore (qualitatively) equivalent 
to flux emergence on the Sun as originally discussed by
Galeev, Rosner \& Vaiana (1979).  As the instability develops, the gas
slides down the expanding loop, and the evacuated loop rises as a
result of the enhanced magnetic buoyancy (the expansion is due to magnetic
pressure although the dynamics are controlled by the down flow due to
gravity along the rising loop). Consequently, as long as the magnetic
pressure at the top of the magnetic loop is larger than the coronal
(radiation, gas and magnetic) pressure, the expansion of the loop continues.
Magnetic buoyancy of an active region thus naturally decreases the
plasma density to the point where the region can efficiently
reconnect. In accretion disk coronae this can take place when either a
new magnetic flux tube emerges into a pre--existing magnetic configuration
or the expansion causes two loops to coalesce; when this happens one
expects at least one current sheet to develop.

If it is assumed that steady reconnection is occurring in such a
current sheet one can solve the energy equation and therefore
determine the temperature of the plasma as a function of the height of
the current sheet in the corona. From this type of analysis it has
been found that, if the sheet rises to a certain height, there will be
no neighbouring equilibrium situation, implying a sudden increase in
temperature (eg. Heyvaerts, Priest \& Rust 1977, Shibata et al. 1989
and references therein). Joule heating can no longer be balanced by
radiation and so the sheet heats up explosively.  The sheet will then
try to achieve a stage where Joule heating can be balanced by thermal
conduction or radiative cooling.  At this point, the current sheet
achieves the critical temperature for the onset of turbulence
(creation of a slow shock), with the result that the flare is
triggered. This is the main phase of the flare, the current sheet
reaches a new steady state, with reconnection based on a marginally
turbulent resistivity (anomalous resistivity) and X-ray flares being
the after effect of the energy release by the slow shocks (e.g. Di
Matteo 1998).  The geometry which we have described here is broadly
consistent with the above picture. In fact, in the soft state the
magnetic loops do not rise high above the accretion disk and the
temperature for the onset of reconnection is expected to be
lower. Current sheets high above the accretion disk, as required for
the hard spectral state, would instead have reached higher
temperatures by the time flares (fast reconnection) are triggered.

\subsection{The two states: different phases of magnetic field dissipation?}
We would now like to incorporate the geometrical description of the
different states with an understanding of the physical reasons which
regulate the selection of the different scale heights for the magnetic
flares.  In the physical context of buoyancy and reconnection, this
implies investigating the reasons why magnetic flux tubes, in the
soft state, are unable to buoyantly expand and grow higher into the
corona before reconnecting and triggering X-ray flares and conversely
why magnetic buoyancy can be much more effective at selecting higher
scale heights in the hard state.  

Let us assume that the rate of magnetic energy dissipation is
constant and the key difference in the two states originates from the way
this energy is released into the corona. 
As already mentioned, magnetic buoyancy instability would take place in
those regions of the disk where shear stresses acting on the a flux
tube generate a magnetic pressure which becomes comparable with (and
eventually exceeds) the disk gas pressure 
The gas pressure
inside a standard (thin) $\alpha$--disk at a distance of $r_{\rm
disk}\simeq 50$ is of the order of $P_{\rm gas}\sim 2\times 10^{12}$
erg cm$^{-3}$ (a viscosity parameters $\alpha_{\rm v}\sim 10^{-2}$ has
been assumed). Magnetic field building up in such a disk will then
become buoyant in those regions where $\beta=P_{\rm gas}/P_{\rm mag}
\approx 1$, i.e. where it reaches a strength $B\sim 6\times 10^6$ G.
Magnetic loops from these regions will eventually emerge into the
coronal regions but their expansion could be halted as soon as they enter
it for reason which we explain below. 
If $B$ in each magnetic loop is set from the buoyancy instability (as
 derived above) and the available magnetic energy is some fraction of
 the accretion power as set by eqn. (1) we deduce that the total
 number of magnetic loops emerging from the disk at any time can be of
 the order of $N_{\rm tot}=10^2-10^3$.  In a very simplified picture we
 can then assume that reconnection takes place every time two such
 loops touch each other. We simulate about $10^3$ loops of radius
 $r=0.3$ (see soft state in Fig. 2) in random positions on the
 surface of an accretion disk of radius $r_{\rm disk} =50$ and find
 that at any given time there will be about 10 loops interacting with
 each other by the time they reach $h \sim 0.3$. This constitutes the number
 of regions $N$ that we have required to be active at any time in the
 soft spectral state (consistently, if there are $N \sim 10$ active regions
 and the timescale for upward buoyancy forces to remove the flux tubes
 from the disk is of the order of $\Omega^{-1} \sim 100 \s$ (Coroniti 1981,
 Stella \& Rosner 1984), where $\Omega$ is the Keplerian angular
 velocity, we expect about $N_{\rm tot} \sim 10^3$ flux tubes to emerge
 from the disk at any given time).


In Section 4 we have shown that, in the hard state, the magnetic field
strength (in the loops) needs to achieve higher values in order to
explain the observed spectrum.  In our scenario, and from the Section
4, this would imply that, in this state, magnetic flux tubes become
buoyant when $B \approxgt$ a few $\times 10^7$~G. This, in turn,
implies that the same total magnetic energy needs to be released in
only $N\sim 10$ `powerful' loops (see eqn. (1)).  These loops rise and
expand high above the disk and, according to our simple simulation,
would start interacting (a reconnection event would take place) when
they reach $h\approx 3$.
The selection of a small number of higher $B$ loops, would explain why
in the hard state flares can be triggered high above the disk.

In this picture, though, it is hard to identify the mechanism that
leads to a change in the spectral states. In other words, it is not
clear why the system would lock up in a certain state and what
physical conditions would cause the transition to another state in the
relevant timescales.  It is interesting to note, though, that the
magnetic field energy density in the powerful loops in the hard state
($B \sim 10^7$ G) is in equipartion with the radiation pressure provided
by the active flaring close to the disk during the soft state ($P_{\rm
rad}\sim 3\times 10^{14} \erg\pcmcu > P_{\rm gas}$). This suggest
that the strong radiation pressure impinging on the disk may be
forcing the magnetic field, in the buoyantly unstable loops, to grow
to larger values in order to be able to rise in the coronal
atmosphere. In other words, we expect that such an intense flare
activity so close to disk, would modify the physical condition of the
disk itself (or at least affect the top layer of the disk).

It is also worth noticing that, according to the picture we have
described above, in both states the magnetic field strength, when
loops reconnect and reach their respective maximum heights, is similar
to the external coronal radiation pressure. It is not clear how the
magnetic field interacts with radiation (either in the corona, or in
the radiation dominated part of the disk) but it seems intriguing that
in the Soft state case, flares are triggered close to the disk and the
ambient radiation pressure (due to intrinsic blackbody emission in
accretion disk) is of the order $P_{\rm rad}
\approx 10^{12} \erg\pcmcu$ in the corona ($B \sim 6 \times 10^6$ G in 
the loops). In the hard state the loops expand as to reach $h \sim 3$
which implies that the magnetic pressure at the top of the expanding
loops ($B\propto h^{-1}$; Shibata et al. 1989) becomes comparable to
the ambient coronal radiative pressure ($P_{\rm rad}\sim 3\times
10^{11} \erg \pcmcu$).  We speculate therefore that the loops
expansion might be halted when the internal magnetic pressure
equalizes the ambient radiation (total) pressure in the
corona. Because of the intrinsically lower $B$ field (see above) in
the soft state magnetic loops would be strongly decelerated as soon as
they enter the corona which explains why no large (high $h$) loops
would be present in this state.

This (admittedly qualitative) picture has some intriguing features,
most notably it accounts for the a priori unpredictable physical
properties required by the spectral observations under reasonable
assumptions.


\section{Variability and reflection}
At least two major observed features need to be accounted for in a
consistent description of the change in spectral states of GBHC,
namely variability and the properties of X--ray reflection features.

In the model discussed above, the extremely rapid and intense
variability on the high energy component observed during hard states
 fit well into the picture. In the situation described above in fact,
during this phase only a few powerful loops would be present at any
time.  Any reconnection event would then produce a significant flux
change and the typical timescale for the formation of a
new structure would probably occur in a detectable timescale of the
order of $h R_{\rm S}/c \sim$ ms. On the contrary the numerous
potentially active region in the soft state would individually be
associated with smaller flux dissipation, occurring over a timescale
$\sim$ 10 $\mu$s, i.e. the variability would result in flickering
not currently detectable.

X-rays impinging on the disk are also expected to be reflected with
different spectral properties in the two states. While in our
situation the covering factor is of the order of $0.5$, the ionization
state of the illuminated disk would be strongly different. In the hard
state, ionization parameters $\xi \approxgt 10^2$ can be inferred over
most of the disk area. In the soft state, $\xi \simeq 10^3$ over the
majority of the disk and $\xi \gg 10^4$ over 10 per cent of 
it.  As mentioned earlier, GBHCs in the hard state have
similar spectra as Seyfert galaxies but with a much less prominent
reflection component.  In the hard state, a ionization parameter $\xi
\approxgt 10^2$,  is in broad agreement reflection parameters deduced
from observations.
Detailed models of reflection for highly ionized disks would need to be
developed in order to make specific predictions for the reflection
features in the hard states of GBHCs.

Nayakshin \& Melia (1997b) have shown that the differences in the
spectrum, in the ionization state of the cold disk and in the iron
lines and edges in the hard state of Cyg X-1 compared to the observed
Seyfert galaxies may be a natural consequence of a change in the
physical state of the 'transition layer' where the hard X-rays from
the active regions are reflected or reprocessed. They also argue that
if X-rays are emitted by localized magnetic flares above the cold
accretion disk, then the energy deposited by the X-rays cannot be
reradiated fast enough to maintain equilibrium, unless the X-ray skin
heats up to the Compton temperature, at which point the gas is mostly
ionized.  As a result, most of the incident X-rays are Compton
reflected back into the active region before reaching the cooler disk
where the reprocessing occurs so that the amount of cooling due to the
soft radiation re-entering the active region is drastically reduced.
 
\section{Discussion}
We have discussed the implications on the spectral states observed in
GBHC of a magnetically--structured corona model, in which magnetic
flares may result from reconnection of flux tubes rising above the
cold accretion disk due to magnetic buoyancy.  Using a simple
geometrical representation which takes into account the scaling of the
different quantities with height of the corona above the disk, we have
shown that the hard X-ray state in GBHC can be produced by flares
located high above the underlying disk via Comptonization of either
the cyclo-synchrotron radiation or the photons produced in the
disk. The soft state is instead due to flaring of active regions close
to it. The soft blackbody component is in this case dominated by the
re-radiated X-ray field and the hard tail is produced via
Comptonization of this same soft radiation.  Depending
on whether the flares are high above the disk or close to it the
spectrum changes from being hard to soft respectively (and therefore
gives rise to the hard state or the soft). When flares are triggered
high above the disk the system is highly photon starved and therefore
the Comtonized spectrum very hard. Conversely when a large numbers of
flares are triggered very close to the disk the the soft photon field
is highly enhanced; this naturally accounts for the a hotter blackbody
component (due to reprocessing in the disk itself) and the very soft
Comptonization spectrum typical of the soft state.
By varying the typical
scale height and scale size of the dissipative regions, the spectra
for all of the five states observed in GX339-4 can be qualitatively
reproduced.  Applying the model to GX339-4 has allowed us to determine
constraints both on the geometrical and physical properties of such
active regions.  

We have qualitatively integrated the geometrical description of the
different states in the physical context of magnetic buoyancy. We have
shown that in the soft state, buoyancy instability in an accretion
disk (which only dissipates a small fraction of the total energy)
naturally favors the emergence of possibly a large number of
relatively weak magnetic field flux tubes (maybe when the dynamo
action is initiated in the disk). Because such tubes have a low
internal pressure their expansion is basically halted as soon as they
enter the coronal regions.  This is the reason why the flaring occurs
really close to the accretion disk itself.  Conversely, the hard state
could be due to a typically different phase of magnetic energy
dissipation (maybe as the dynamo action builds higher magnetic
fields).  This would take place in a small number of very intense
magnetic field loops.  The reason why flares would reach higher scale
heights, in this model, can then be ascribed to the intrinsically more
intense magnetic field loops which would lead the loops to grow and
expand before efficient reconnection takes place. Also, large loops
would naturally form from small scale ones. Reconnection between small
loops can drive a time--dependent inverse cascade process which leads
to the formation of larger structures and cause spectral changes.


While the model described is certainly too schematic both in its
geometrical/uniformity and physical assumptions it seems to 
account qualitatively for other observed properties of GBHC, namely
the variability and characteristics of the reflected X--ray component.

In other models proposed to account for the spectral states of black
hole candidates the amount of power dissipated in the accretion disc
varies greatly from one state to the other, so that it is almost
negligible in the hard state while dominating in the soft one. The
model examined here on the contrary postulates that a constant
fraction of the total power is radiatively dissipated in the
disk. Most of the energy is instead released in magnetic coronal
flares at different heights above the disk. The amount of heating and
re--radiation from the accretion disk itself during the triggering of
flares can naturally give rise to all of the spectral components
observed in the different states of the source.

\section{Acknowledgments}
Trinity College (TDM), the Italian MURST and IOA (PPARC theory grant)
(AC) and the Royal Society (ACF) are thanked for financial support.

\end{document}